\documentclass[twocolumn,showpacs,amsmath,amssymb]{revtex4}

\usepackage{graphicx}
\usepackage{dcolumn}
\usepackage{bm}

\begin{document}
\title{Heating of solar chromosphere by electromagnetic wave absorption in a plasma slab model}
\author{D. Tsiklauri}
\author{R. Pechhacker}
\affiliation{Astronomy Unit, School of Mathematical Sciences, 
Queen Mary University of London, Mile End Road, London, E1 4NS, United Kingdom}
\date{\today}
\begin{abstract}
The heating of solar chromospheric inter-network regions by means of the absorption of electromagnetic (EM) waves
that originate from the photospheric blackbody radiation is studied in the framework of a plasma slab model.
The absorption is provided by the electron-neutral collisions in which electrons oscillate in the EM wave
field and electron-neutral collisions damp the EM wave.
Given the uncertain nature of the collision cross-section due to the plasma micro-turbulence,
it is shown that for plausible physical parameters, the 
heating flux produced by the absorption of EM waves in the chromosphere is between $20 - 45$ \%
of the chromospheric radiative loss flux requirement. It is also established that there is an optimal
value for the collision cross-section, $5 \times 10^{-18}$ m$^{2}$, that produces the maximal 
heating flux of 1990 W m$^{-2}$.
\end{abstract}	


\maketitle

\section{Introduction}

The problem of heating of the solar atmosphere, the sharp temperature raise from photospheric 6000K to few $10^6$K 
in the corona has been a long standing problem of solar physics. There is no lack of possible heating
mechanisms in the {\it corona} with the two main candidates being so-called direct current (DC) models that are based
on the magnetic reconnection and alternating current (AC) models that are based on magnetohydrodynamic
(MHD) wave dissipation \cite{2005psci.book.....A}, alongside with few dozen other, less widely 
accepted \cite{1994ApJ...427..446S,2006A&A...455.1073T}
or less successful ones \cite{2005A&A...441.1177T}. 
In the {\it chromosphere}, however, until recently, a general agreement was that it is heated by the absorption of
the acoustic waves. A distinction needs to be drawn between wave heating of different parts of the chromosphere.
The chromosphere of the quiet Sun can be broadly split into two parts: (i) the magnetic network, which marks 
the boundaries in between the super-granulation cells; and (ii) the inter-network regions, which constitute the 
bulk surface area of the chromosphere (i.e. the super-granulation cell interiors).

In the magnetic network the magnetic field is nearly radial (vertical) and quite strong (of the order of few kG).
Since the strong magnetic field there provides a substantial amount of free energy, in principle it seems reasonable to believe
that the magnetic network can be heated by the dissipation of MHD waves \cite{2008ApJ...680.1542H}. However, the role of
magnetic reconnection in the heating of the magnetic network cannot be discounted. For example, there seems to be an
evidence of forced magnetic reconnection taking place in the photosphere \cite{2010ApJ...712L.111J}, 
as well as in chromosphere \cite{2010ApJ...713L...6C}. The estimates of the heating flux
produced by plausible reconnection models seem to fall short by up to two orders of magnitude
from the quiet chromosphere and coronal heating requirements \cite{1999ApJ...524..483L}.
At the same time  Ref.\cite{2008ApJ...680.1542H} and other similar works, that base their
conclusions on the numerical simulation results, do not make precise predictions for the heating rates 
produced by the dissipation of MHD waves in the chromospheric magnetic network.

The situation with the heating of inter-network regions where magnetic fields are weak
is even worse than with the magnetic network. On one hand, this is because of the lack 
of the magnetic energy. On the other hand, the results of Ref.\cite{2005Natur.435..919F} 
indicate that the acoustic energy flux of the 
high-frequency (10-50mHz) acoustic waves (that were previously believed to constitute the 
dominant heating mechanism of the chromosphere) falls short, by a factor of at least ten, 
to balance the radiative losses in the solar chromosphere. This led them to a conclusion that the
acoustic waves cannot constitute the dominant heating 
mechanism of the solar chromosphere. This conclusion has been challenged by Ref.\cite{2008JApA...29..163K}, who
suggests that the observations reported by Ref.\cite{2005Natur.435..919F}  only detect 10\% of the
acoustic wave flux perhaps because of the limited spatial resolution.
Ref.\cite{2010ApJ...723L.134B} report somewhat higher than usual 
flux carried by the acoustic waves at photospheric heights (250 km) as well as compile
a useful list of up to date acoustic heating flux measurements.

In this context, in this paper we explore an alternative to the acoustic wave heating idea
of the quiet chromosphere. In particular, we investigate the following
proposition: 

(i) It is known that the solar irradiance spectrum, that comes out of photosphere,
is well approximated by an effective blackbody at a temperature of
$T=5762$ K, in the frequency range of $f =30-1667$ THz (corresponding to the 
wavelengths range of $10 - 0.18$ $\mu$m) (see e.g. Figure 2.3 from
Ref.\cite{2005psci.book.....A}). Therefore, we assume that the radiative 
heating flux with the Planckian brightness distribution as a function of frequency
penetrates the lower part of the solar atmosphere (photosphere, $h=0 - 500$ km and chromosphere, $h=500-2200$ km).

(ii) Instead of solving radiative transfer equations, we take the photospheric blackbody flux of  $T=5762$ K, and
quantify how much electromagnetic (EM) radiative flux is absorbed using a plausible model
for EM wave absorption, which is based on Ref.\cite{2003ITPS...31..405T} plasma slab model combined 
with the VAL-C  model of 
chromosphere \cite{1981ApJS...45..635V}. Ref.\cite{2003ITPS...31..405T} plasma slab model is based on splitting
a smoothly varying,  non-uniform density, weakly ionised plasma with the uniform magnetic field along the density gradient,
into a set of thin sub-slabs with a uniform density in each sub-slab - thus providing a discretized version
of the smooth density profile. The absorption of the EM radiation is based on two
physical effects: electron-neutral collisions and electron cyclotron resonance.
For the considered in our model radial magnetic field value of 0.2 kG, electron cyclotron frequency is
$f_{ce}=eB/2 \pi m_e= 0.00056$ THz. 
Also, for the considered model parameters, the ratio of electron-neutral collision frequency and 
electron cyclotron frequency, $\nu_{en}/f_{ce} << 1$, which ensures that the collisions would not
affect any electron cyclotron resonance damping.
Therefore EM wave absorption via  electron cyclotron resonance
 is negligibly small in the considered range of frequencies $2 - 2000$ THz.
 We refer the interested reader to Ref.\cite{2003ITPS...31..405T} for the details
 of the plasma slab model. However, we re-iterate the key points of the model in Section 2.

As a result we find that for plausible physical parameters, the 
heating flux produced by the absorption of EM waves in the chromosphere is between $20 - 45$ \%
of the VAL-C radiative loss flux requirement. We also establish that there is an optimal
value for the collision cross-section, $5 \times 10^{-18}$ m$^{2}$, that produces the maximal 
heating flux of 1990 W m$^{-2}$.

The paper is organised as follows: In Section 2 we describe the model. In Section 3 we present the numerical
results and we close with the Conclusions in Section 4.

\section{The model}

The interaction of EM wave with a plasma slab has been a subject of a number of studies see e.g. Ref.\cite{2003ITPS...31..405T},
references therein, and a more recent work \cite{pop2}.

Following the general approach of Ref.\cite{2003ITPS...31..405T}, we consider a slab that contains 
22 sub-slabs each having thickness of 100 km. 
In each sub-slab density and temperature are assumed to be uniform, but these vary as we go from one slab to the next.
This variation is prescribed by the VAL-C model \cite{1981ApJS...45..635V}. This way, 
in the first slab, that corresponds to the 50 km above photospheric level, we have temperature of 
T=5840 K, neutral hydrogen number density $n_H=9.203\times 10^{22}$ m$^{-3}$, 
electron number density of $n_e=2.122\times 10^{19}$ m$^{-3}$.
In the final 22nd slab, that corresponds to the 2200 km above photospheric level
the plasma parameters are  T=24000 K, $n_H=1.932\times 10^{16}$ m$^{-3}$, 
$n_e=2.009\times 10^{16}$ m$^{-3}$. In between these values we use linear interpolation
for the thermodynamic parameters. The uniform magnetic field with $B=0.02$ T (0.2 kG) 
is directed through all sub-slabs 
in the vertical (radial) direction. We use a relatively small value of $B$ that is commensurate with
the chromospheric inter-network regions. 
However, we note that the model outcomes (such as e.g. the resultant EM wave absorption) 
depend weakly on the magnetic field value used (at least in the plausible
range of variation of the field in the chromosphere).
Also the fact that the magnetic field is vertical describes
the quiet Sun reasonably well. Ref.\cite{2010A&A...517A..37S} found no evidence for any 
predominance of horizontal fields
on the quiet Sun. They find that (i) the angular distribution of the field varies
steeply with flux density. (ii) For the largest flux densities the distribution
is extremely peaked around the vertical direction. (iii) For the smaller vertical flux densities 
the distribution widens to
become asymptotically isotropic in the limit of zero flux density.
The apparent dominance of horizontal fields for flux densities
below 5 G is shown to be an artifact of noise.

Our representation of the number density and temperature variation in the photosphere and chromosphere
by means of uniform sub-slabs is valid when the variation over height is slow. This is indeed the case
in the photosphere and chromosphere (we exclude the transition region and corona from our consideration also
on the grounds that the dispersion relation that we use below is applicable for {\it weakly} ionised plasma). 
The expression for the complex dielectric constant for the weakly ionised, magnetised plasma with 
the angle between the propagation direction of the incident EM wave and the magnetic field $B$,
$\theta=0$ is
taken from  Ref.\cite{2003ITPS...31..405T}:
\begin{equation}
\tilde \varepsilon_r = 1 -\frac{\omega_{pe}^2 / \omega^2}{\left[1 - i \nu_{en}/\omega \right] \pm \omega_{ce}/ \omega}.
\end{equation}
Here $\omega_{pe}=\sqrt{n_e e^2/(\varepsilon_0 m_e)}$, $\omega_{ce}=eB/m_e$ and $\nu_{en}$ (see Eq.(7) below for the definition) 
are the electron plasma frequency, electron cyclotron angular frequency and
electron-neutral collision frequency, respectively. 
The angular frequency $\omega$ of the EM harmonic is related to the frequency, $f$, in the usual way $\omega = 2 \pi f$.
The sign $\pm$ refers to the left- and right-hand polarisation of the
EM wave and $i$ is the imaginary unit.
The EM wave reflection coefficient at the $(k+1)$th sub-slab interface is given by
\begin{equation}
\Gamma (k+1) =\frac{\sqrt{\tilde \varepsilon_r(k)} - \sqrt{\tilde \varepsilon_r(k+1)}}{\sqrt{\tilde \varepsilon_r(k)} + \sqrt{\tilde \varepsilon_r(k+1)}}.
\end{equation}
One can calculate the reflected and transmitted power, $P_r$ and $P_t$, respectively, using formulae:
\begin{eqnarray}
P_r(f) &=& P_i(f)\Biggl[|\Gamma(1)|^2 + \nonumber \\
&&\sum_{j=2}^{22}\left(|\Gamma(j)|^2 
\prod_{i=1}^{j-1}{ \left[ e^{-4 \alpha (i) d}\left(1- |\Gamma(i)|^2\right) \right]} \right) \Biggr]
\end{eqnarray}
\begin{equation}
P_t(f) = P_i(f) 
\prod_{i=1}^{22}{ \left[ e^{-2 \alpha (i) d}\left(1- |\Gamma(i)|^2\right) \right]}, 
\end{equation}
where $P_i$ is the incident EM wave power, $d$ is the sub-slab thickness and 
$\alpha (i)$ is the real part of the complex propagation constant of a plane wave in a magnetised  plasma corresponding to slab $i$
\begin{equation}
\alpha (i) =\frac{\omega}{c} {\rm Re} \left[\sqrt{- \tilde\varepsilon_r(i)} \right]
=\frac{\omega}{c} \sqrt{ \frac{ | \tilde\varepsilon_r(i) | - {\rm Re} [ \tilde\varepsilon_r(i) ] }{2}}.
\end{equation}
Note that Eqs.(1)-(4) are identical to the corresponding equations from Ref.\cite{2003ITPS...31..405T}.
As in Ref.\cite{2003ITPS...31..405T} we neglect the effect of multiple reflections of waves, i.e. each wave gets reflected 
at a maximum of one slab interface, however may be absorbed in any location.

The total absorbed power, $P_a$ is then given by
\begin{equation}
P_a(f) = P_i(f) -P_r(f) -P_t(f) .
\end{equation}
Note that for $\theta=0$ the above expressions for the reflection, transmission and absorption coefficients 
are independent of the EM wave polarisation, except for 
the $\pm$ sign in Eq.(1) for the left- and right-hand polarisation of the
EM wave (see e.g. Ref.\cite{hw78}, pp. 71-94). 
A simple numerical code was written to calculate the above absorbed, reflected and transmitted EM power.
We have tested the code by successfully reproducing Figures 1-9 from Ref.\cite{2003ITPS...31..405T}, using their set of 
physical parameters.

The above formulae contain the electron-neutral collision frequency $\nu_{en}$. 
For the latter we use the standard expression
from Ref.\cite{nrl}, p.39, 
\begin{equation}
\nu_{en} = n_0 \sigma \sqrt{k T_e /m_e}, 
\end{equation}
where $n_0$ is the neutral number density, $\sigma$ is the collision cross-section and the square root is essentially
the electron thermal speed. For $n_0$ we use values of neutral hydrogen number density $n_H$ in each sub-slab, and for $T_e$ we use
the temperature $T$; both according to the VAL-C model.  Thus, in each presented numerical model run we regard $\sigma$ as fixed. However, since
$\nu_{en}$ is a function of density and temperature, each  sub-slab has its own set value. 

In order to calculate the absorbed EM flux we use 
equal $1/2$ statistical weights of the left- and right-hand polarised EM wave total absorbed powers $P_{a,L}$ and $P_{a,R}$
as following
\begin{equation}
P_a(f) = \frac{1}{2} P_{a,L}(f) + \frac{1}{2} P_{a,R}(f).
\end{equation}
Thus, the total absorbed EM flux can be calculated using
the following integral
\begin{equation}
F_a [\rm  W m^2]= \pi \int_{f_{min}}^{f_{max}} A(f) B_f(f) df,
\end{equation}
where $B_f(f)$ is the Planck function $B_f(f)=2 h f^3 / \left[c^2 (\exp[h f/(k T)] -1) \right]$ and 
$A(f)$ is the total absorption coefficient as a function of frequency given by the following expression 
\begin{equation}
P_a(f) = P_i(f) A(f).
\end{equation}

\section{Results}

The dissipation coefficients in the solar atmosphere are not known precisely.
There are good reasons to believe that the dissipation coefficients, such as resistivity and viscosity,
which in turn depend on plasma species mutual collision cross-sections and collision frequencies 
(including $\nu_{en}$), have so-called
"anomalous" values. The "anomaly" is in the sense of their departure (mostly increase) from the classical (laminar) plasma
transport theory values. It is believed that the anomalous dissipation coefficients result from the plasma
micro-turbulence. The latter provides additional centres of scattering for the plasma particles
in addition to their mutual collisions. 
In effect this means that in Eq.(7) we should use $\sigma =\sigma_{en} + \sigma_{turbulent}$, i.e.
the effective anomalous cross-section is a sum of usual electron-neutral and turbulent plasma cross-sections.
Note that $\sigma_{en} \approx 5 \times 10^{-19}$ m$^2$ and is 
weakly dependent on temperature \cite{nrl}. Little is known about the $\sigma_{turbulent}$ apart from it can be
orders of magnitude larger than $\sigma_{en}$.
Since the theory of plasma micro-turbulence
is  not complete, we simply vary the collision cross-section $\sigma$ in Eq.(7)
by a few orders of magnitude, in order to study the effects of anomalous dissipation coefficients
on the quiet chromosphere heating. We start presentation of the results by plotting, in Figure 1, 
the EM wave absorption 
coefficient, $A(f)$, as a function of frequency for various values of $\sigma$.
The considered frequency range is commensurate with the range $f =30-1667$ THz where 
the solar irradiance spectrum, that comes out of photosphere,
is well approximated by an effective blackbody at $T=5762$ K (see e.g. Figure 2.3 from
Ref.\cite{2005psci.book.....A}).

\begin{figure}    
   \centerline{\includegraphics[angle=-90,width=0.5\textwidth,clip=]{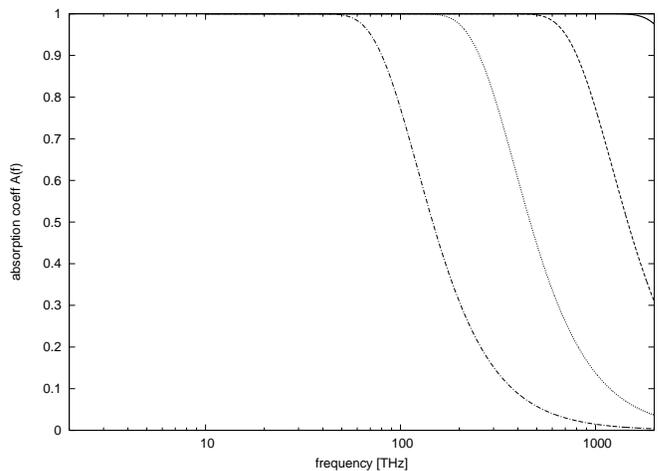}
              }
\caption{Absorption coefficient $A(f)$ is a function of EM wave frequency $f$ for different 
values of the cross-section $\sigma$. Solid line corresponds to $\sigma=5 \times 10^{-16}$, 
dashed $\sigma=5 \times 10^{-17}$, dotted $\sigma=5 \times 10^{-18}$, 
dot-dashed $\sigma=5 \times 10^{-19}$.}
\end{figure}
We gather from Figure 1  that (i) a decrease in $\sigma$ generally results in an overall reduction of the
absorption coefficient; and (ii)  different frequencies are absorbed differently -
high frequencies show weaker absorption than low ones. As to the observation (i), 
the explanation can be provided by analysing by Eqs. (1) - (4). 
It is clear from Eqs. (3) and (4) that the electron-neutral collision 
frequency $\nu_{en}$ provides the only dissipation effect. 
The exponential factor $e^{-\alpha d}$ dominates the heat flux absorption. 
For the considered model parameter range,
it can be shown by calculating $\partial \alpha / \partial \nu_{en}$ 
derivative numerically that
$\partial \alpha / \partial \nu_{en} > 0$ .
Therefore one can deduce that the heat flux absorption decreases with the decrease in cross-section.
As to the observation (ii), a possible physical explanation could be that when the frequency of the
EM radiation is increased, electrons due to their small but 
finite inertia do not have time to catch up with (i.e. couple to) the EM wave. Hence the EM wave frequency increase
results in a decrease of the absorption coefficient.

In Figure 2 we show the behaviour of the integrand of Eq.(9), 
$\pi A(f) B_f(f)$, which has a physical meaning of the
absorbed EM flux density (i.e. flux per unit frequency)
as a function of frequency 
for various values of $\sigma$.
\begin{figure}    
   \centerline{\includegraphics[angle=-90,width=0.5\textwidth,clip=]{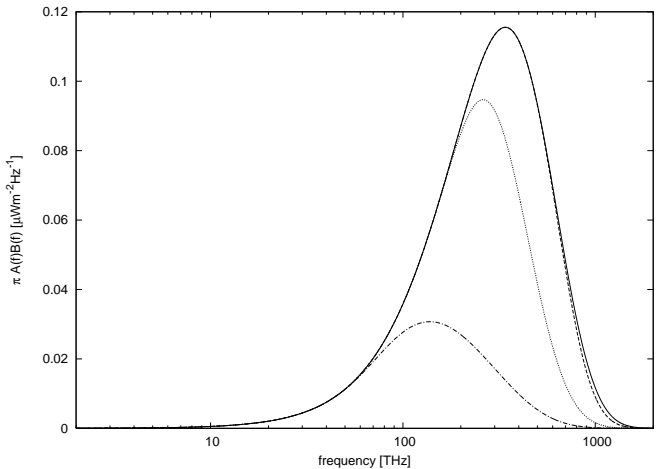}
              }
\caption{The absorbed EM flux density, $\pi A(f) B_f(f)$,  as a function of frequency 
for various values of $\sigma$. 
Solid line corresponds to $\sigma=5 \times 10^{-16}$, dashed $\sigma=5 \times 10^{-17}$, 
dotted $\sigma=5 \times 10^{-18}$, dot-dashed $\sigma=5 \times 10^{-19}$.}
      \end{figure}
We observe that due to the frequency dependence of $B_f(f)$  that has a peak,
the absorbed EM flux density is also peaked. The two conclusions that follow are:
 the decrease in the cross-section results in (i) the overall reduction of the  
the absorbed EM flux density; and (ii) shift of the absorbed EM flux density's peak
towards the smaller frequencies.

Next we calculate the volumetric heating rate produced by the absorption of the EM waves. 
Above we have presented the behaviour of the total absorption coefficient and the absorbed EM flux density for the full set of 22 slabs 
as a function of frequency. We obtained the expression for the total absorbed flux by integrating over 
the relevant interval of frequencies using Eq.(9). In order to obtain the distribution of the
absorbed EM energy as a function of height, we perform a numerical differentiation of the calculated total absorbed flux
values. The differentiation yields the heating rate $H(k)$ in slab $k$,
\begin{equation}
H(k) = \frac{F_a(k) - F_a(k-1)}{d}
\end{equation}
where the $F_a(k)$ are calculated by ignoring the existence of slabs above slab number $k$.

In Figure 3 we plot the heating rate for the different values of $\sigma$ as a function of height above photosphere.
Clearly, the model predictions for the heating rate fall short in matching the 
empirical radiative loss calculated by the VAL-C model of the chromosphere. We gather that
(i) the heating rate decreases rapidly with  height; and (ii) in the region of 500 km - 2200 km above photosphere 
there is an optimal value for $\sigma$ that produces a maximal heating rate. The latter is evidenced by the
fact that $\sigma=5 \times 10^{-18}$ (dotted) curve is above the both $\sigma=5 \times 10^{-17}$ (long-dashed)
and $\sigma=5 \times 10^{-19}$ (dot-dashed) curves.     
\begin{figure}    
   \centerline{\includegraphics[angle=-90,width=0.5\textwidth,clip=]{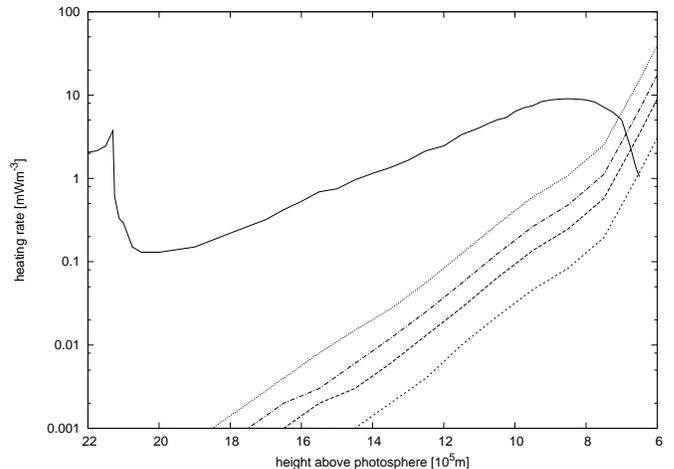}
              }
\caption{The heating rate for the different values of $\sigma$ as a function of height above photosphere. 
The VAL-C empirical model (solid); the next four curves are calculated using Eq.(11):
$\sigma=5 \times 10^{-17}$ (long-dashed), 
$\sigma=5 \times 10^{-18}$ (dotted), $\sigma=5 \times 10^{-19}$ (dot-dashed), 
$\sigma=5 \times 10^{-20}$ (short-dashed).}
      \end{figure}
Note that the model predictions for the heating rate presented in Figure 3 
underestimate the actual heating rate produced by the absorption of EM waves.
This is due to the following reasons: (i) an  error introduced by the numerical
differentiation with respect to height (it is known that the numerical differentiation
always introduces the numerical diffusion); and (ii) ignoring the contribution to
the absorption due to reflected waves. 
The former is unavoidable, while the latter can be regarded as a shortcoming of 
our heating rate calculation. Our motivation for calculating the (volumetric) heating rate
as a function of height
was to compare our model predictions to the results of VAL-C empirical radiative cooling rate 
(see their Figure 49). Naturally, we can improve our calculation by {\it directly}
calculating the total heating flux of the chromosphere. We do this by applying
Eq.(9) for all 22 sub-slabs that cover entire the photosphere and chromosphere (heights of $0-2200$ km) and
then subtracting contribution from the first 5 photospheric sub-slabs ($0-500$ km), because we are concerned only with
the chromosphere.
Such {\it heating flux} [W m$^{-2}$] calculation is 
free from the above two sources of the underestimation that arose in the calculation of 
the {\it (volumetric) heating rate} [W m$^{-3}$]. Further, 
the most exact calculation of the total absorbed EM flux is as follows:

In order to extract the total absorbed flux of a slab interval $[a;b]$ , we have to account for the possibility of waves being reflected at an interface higher up than slab $b$ and 
being absorbed within the interval on their way back. In the following we will use the notation $A|_a^b$ for 
the absorption coefficient of a plasma slab bounded by slabs number $a$ and $b$, both included in the interval.
Further we define
\begin{equation}
A|_a^b = 1 - R|_a^b - T|_a^b
\end{equation}
where $R|_a^b$ and $T|_a^b$ represent reflection and transmission coefficients of the interval and can be calculated by Eq.(3)-(4). 
At this point we stress that by using Eq.(3)-(4) we neglect the effect of multiple reflections of waves, meaning that each wave gets reflected 
at a maximum of one slab interface, however may be absorbed in any location. Further we define that only coefficients that follow $1 \leq a \leq b \leq N$, 
where $N$ is the total number of slabs, are non-zero.
First we consider the case $a=1$ and $b$ arbitrary. We write for the absorption coefficient
\begin{equation}
A|_1^b = A|_1^N - A|_{b+1}^N T|_1^b
\end{equation}
where the last term on the right hand side accounts for absorption of waves within layers higher up than slab $b$ that have passed 
through the interval. Extending the formalism to arbitrary $a$ it holds true that
\begin{equation}
A|_a^b = A|_1^N - A|_{b+1}^N T|_1^b - A|_1^{a-1} - T|_1^{a-1} R|_a^N (A|_1^{a-1})^*
\end{equation}
where
\begin{equation}
(A|_1^{a-1})^* = 1 - T|_1^{a-1} = 1 - \prod_{i=1}^{a-1}{ \left[ e^{-2 \alpha (i) d} \right]}.
\end{equation}
The factor $(A|_1^{a-1})^*$ accounts for the 
absorbed radiation within the interval $[1;a-1]$ but neglects the multiple reflections.
We gather from Eq.(14) that we have added a term that gives absorptions at layers lower than $a$ and another term that accounts for 
waves that have initially been transmitted through those lower layers, then get reflected at higher levels and absorbed below level $a$ 
on their way back down.

The results of the these calculations, inserting $N=22$, $a=6$ and $b=22$, are shown in Figure 4,
where we plot the total absorbed EM flux as a function of $\sigma$.   
\begin{figure}    
   \centerline{\includegraphics[angle=-90,width=0.5\textwidth,clip=]{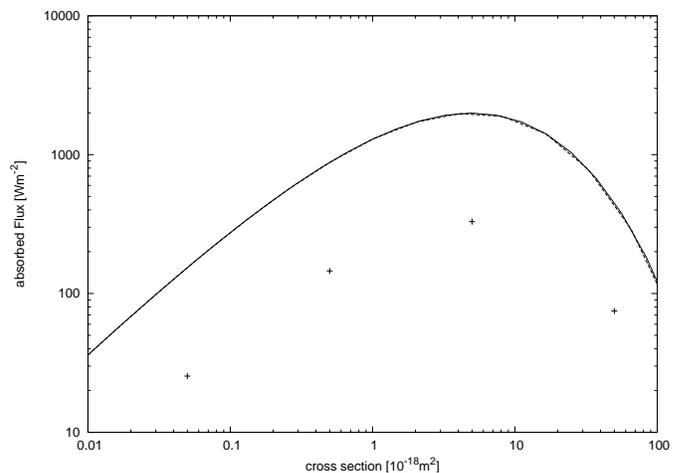}
              }
\caption{The total absorbed EM flux as a function of $\sigma$. 
The solid curve corresponds to the precise calculation according to Eq.(14).
The dashed curve corresponds to the 
the calculation by applying
Eq.(9) for all 22 sub-slabs that cover entire photosphere and chromosphere (heights of $0-2200$ km) and
then subtracting the contribution from the first 5 photospheric sub-slabs ($0-500$ km).
Four crosses correspond to the area under the long-dashed, dotted, dot-dashed and short-dashed
curves in Figure 3 that cover chromospheric heights between 6th and the 22nd sub-slabs $500-2200$ km.}
      \end{figure}
We gather from Figure 4 that (i) there is an optimal
value for $\sigma = 5 \times 10^{-18}$ m$^{2}$ that produces the maximal heating flux of 1990 W m$^{-2}$.
The latter is about $45$ \% of the chromospheric radiative losses; 
(ii) For the value of
$\sigma = 5 \times 10^{-19}$ m$^2$ that is predicted by the classical (laminar) plasma
transport theory \cite{nrl}, which ignores contribution from the plasma-microturbulent transport,
the total heating flux by the absorption of EM waves in the chromosphere our model predicts
880 W m$^{-2}$. Which is only 20\% percent of the heating flux
requirement of 4280 W m$^{-2}$ \cite{1981ApJS...45..635V}.
Since precise value of $\sigma$ is unknown due to a lack of the 
detailed plasma-microturbulent transport theory, we conclude that the actual
heating flux produced by the absorption of EM waves in the chromosphere is between $20 - 45$ \%
of the VAL-C radiative loss flux requirement.
Note that the behaviour described in Figure 4 that the heating flux has a maximum
when plotted as a function of cross-section $\sigma$ is true when only chromospheric 
sub-slabs $6-22$ are considered (i.e. when contribution from the photospheric 
sub-slabs $1-5$ is not included). We also remark that the difference between the
solid curve that corresponds to the precise calculation according to Eq.(14) and
the dashed curve that corresponds to the 
the calculation by applying
Eq.(9) for all 22 sub-slabs that cover entire photosphere and chromosphere (heights of $0-2200$ km) and
then subtracting the contribution from the first 5 photospheric sub-slabs ($0-500$ km) is
rather small and barely distinguishable to the plotting accuracy.
The close overlap of the two curves is an indication of the last term in Eq.(14) being 
negligibly small. This term accounts for the radiation that is initially transmitted into 
slabs 6 and higher, reflected somewhere in the higher slabs and then again absorbed 
within the slabs 1-5. It is clear from Fig. 3 that the absorption in 
the lower layers is higher compared to the upper ones. Hence, the  
last term in Eq.(14) will not make a considerable contribution when the reflection 
and transmission coefficients $R|_a^N$ and $T|_1^{a-1}$ are sufficiently small, such 
that the overall product $T|_1^{a-1} R|_a^N (A|_1^{a-1})^*$ is negligible.  
For for the chromospheric parameters the last term in Eq.(14) is negligibly small,
thus the plotted curves in figure 4 are barely distinguishable.

\section{Conclusions}

In this paper we put to test a simple proposition that some of the problems of the
chromospheric inter-network regions (regions with weak magnetic field which constitute the
bulk of solar chromosphere surface), discussed in the Introduction section can be alleviated by
inclusion of the absorption of photospheric EM radiation in the plasma sub-slab based model.

On one hand, we know that that the solar irradiance spectrum, that comes out of photosphere,
is well approximated by an effective blackbody at a temperature of
$T=5762$ K, in the frequency range of $f =30-1667$ THz. Therefore, we can assume that the radiative 
heating flux with the Planckian brightness distribution as a function of frequency
illuminates lower part of the solar atmosphere (photosphere, $h=0 - 500$ km and chromosphere, $h=500-2200$ km).

On the other hand, instead of solving radiative transfer equations, we can take photospheric blackbody flux of  $T=5762$ K, and
quantify how much electromagnetic (EM) radiative flux is absorbed using a plausible model
for the EM wave absorption which is based on Ref.\cite{2003ITPS...31..405T} plasma slab model combined 
with VAL-C  model of 
chromosphere \cite{1981ApJS...45..635V}. Our model is based on splitting
a weakly ionised plasma slab with the uniform magnetic field along a smooth density gradient,
into a set of narrow sub-slabs with a uniform density in each slab - hence providing a discretized version
of the smooth density gradient. In the relevant frequency range ($2 - 2000$ THz), the absorption of the 
EM radiation is due to the electron-neutral collisions, while the electron cyclotron resonance can be ignored.
The absorption of the EM radiation due to the electron-neutral collisions happens because the 
electrons oscillate in the EM wave field and electron-neutral collisions (i.e. their mutual friction) 
then results in the damping of the EM wave.

We also include a contribution to the cross-section from the anomalous plasma micro-turbulence, which we incorporate
in an additive way.

Our model has two potential weaknesses: (i) to what extent the radiative 
heating flux of the photosphere deviates from the Planckian brightness distribution as a function of frequency?
and (ii) whether the {\it absorption} of EM radiation can be described by the plasma sub-slab model which assumes
local thermodynamic equilibrium (LTE)?
Concerning the first question we state that it is well known that the solar irradiance spectrum, that comes out of photosphere,
is well approximated by an effective blackbody at a temperature of
$T=5762$ K, in the frequency range of $f =30-1667$ THz. Also, there are plausible models of solar photosphere
that use LTE assumption (see e.g. Ref.\cite{2008A&A...486..951G}) that implies the applicability of the  Planckian 
brightness function. As to the second question we remark that indeed in order to properly work out the
chromospheric (radiative) {\it losses}, one needs to consider non-LTE effects. Strong radiation field in the
chromosphere can drastically alter the occupation numbers of the energy levels in atoms, thus producing
non-LTE net radiative loss \cite{1978stat.book.....M}. We assert, however, that the chromospheric {\it heating} process can 
be regarded as an LTE process,
given the {\it steady} inflow of EM radiation from the photosphere. After all, the chromospheric heating models
that are based on the absorption of acoustic shock wave energy use equations of hydrodynamics which imply
LTE, and it is only the radiative {\it losses} that are treated by the non-LTE radiative transport as in e.g. 
Ref.\cite{2005Natur.435..919F}.

As a result we find that:

(i) for plausible physical parameters, the 
heating flux produced by the absorption of EM waves in the chromosphere is between $20 - 45$ \%
of the VAL-C model radiative loss flux requirement. 
The variation range is because of the uncertainties in the collision cross-section 
due to the plasma micro-turbulence.

(ii) We also established that for absorption in the region 500 km - 2200 km above photosphere there is an optimal
value for $\sigma = 5 \times 10^{-18}$ m$^{2}$ that produces the maximal heating flux of 1990 W m$^{-2}$.

From the observational point of view, if the absorption of 
EM waves in the frequency range $2 - 2000$ THz has a 
significant role to play in the heating of quiet chromosphere, 
as suggested by our findings, then the plasma slab model also predicts that:

(i) There is a good case for the electron-neutral 
anomalous collision cross-section to be a factor of 10 larger 
than the value predicted by the classical plasma transport.
Ref.\cite{2003PhPl...10..319T} presented plasma resistivity 
(which is proportional to both the collision frequency and 
cross-section)
measurements in
the reconnection current sheet of the Magnetic Reconnection
Experiment. They established that in some regimes, the
measured resistivity values can be more than an order of
magnitude larger than the classical Spitzer value. 
Therefore it would seem likely that 
the collision cross-section in the chromosphere would also
assume some anomalous value. 

(ii) There should be a good correlation of the total solar 
irradiance with the Mg-index (which represents the
chromospheric excess radiation relative to the photosphere) on a
long-term (1 month or more) timescale. This is because our model
takes photospheric blackbody EM wave flux as the source of energy,
that irradiates chromosphere from below (a torch shining from the below analogy
is relevant here).
In fact, this is exactly what is observed: Ref.\cite{2007ASPC..368..481S}
presents the data that shows that the total solar 
irradiance and the Mg-index have a correlation coefficient of 0.8
using monthly data averages (see their Figure 1 and pertinent discussion).
The correlation is somewhat worse of a shorter timescales, e.g.
daily data averages -- this has a good explanation in that
contribution from 
the solar features such as sunspots and faculae (that affect photospheric total
solar irradiance) and plages (that affect chromospheric brightness and are 
in fact mapped closely to the faculae below)
average out on the long timescales and generally track to solar
activity cycle (that has a proxy of number of sunspots).

(iii) Unlike in the photosphere, the chromospheric brightness
should not decrease with the increase of the magnetic field.
This is because in our model EM wave absorption depends on the
magnetic field rather weakly. This is also what is observed: 
Ref.\cite{1989ApJ...337..964S} find that CaII K line core contrast
(the relative difference between the intensity at a given
magnetogram and the quiet Sun intensity) that is a measure of
chromospheric brightness is weakly increasing with the magnetic field as
$\propto B^{0.6}$. In the photosphere the contrast of a continuum in the
green part of the solar spectrum initially increases with $B$ up to 0.02 Tesla
but then sharply decreases with $B$ above 0.05 Tesla.
 Ref.\cite{2007ASPC..368..481S} explain chromospheric
rise of brightness with the increase of magnetic field
for small $B$'s ($0.01$ T) is due to the density increase of the magnetic
flux tubes, and for large $B$'s ($> 0.05$T) subsequent slower rise is due to the 
quenching of the wave activity.
As Ref.\cite{1989ApJ...336..475T} have shown, the strong magnetic
fields inhibit average horizontal flow speeds in the granules. 

Thus in conclusion the plasma slab model predictions seem also to conform
with the available observational data.

\begin{acknowledgments}
The authors are financially supported by the HEFCE-funded 
South East Physics Network (SEPNET). The authors would like to thank an 
anonymous referee whose comments contributed to an improvement of this paper. 
\end{acknowledgments}


\end{document}